# Nonlinear Modeling of a PEM Fuel Cell System; a Practical Study with Experimental Validation

Seyed Mehdi RakhtAla[1*] and Roja Eini[2]

[1]Department of Electrical Engineering, Golestan University, Gorgan, Iran.

[2]Department of Electrical and Computer Engineering, Babol University of Technology, Babol, Iran.

*Corresponding Author's E-mail: *sm.rakhtala@gu.ac.ir*

**Abstract**

   In this paper, a nonlinear six order model is proposed for a proton exchange membrane fuel cell (PEMFC) as a control-oriented electrochemical model. Its validation is performed on a specific single cell PEMFC with effective dimension of 5 cm×5 cm. This model is described in the nonlinear state space form with 6 state variables. Load current and DC voltage are considered as measurable disturbance and control input respectively. Besides, the model includes fuel cell stack and its auxiliary components as well. In this survey, a nonlinear state space representation is derived by arranging nonlinear equations and combining them with auxiliary components model. The proposed model can be successfully used to design nonlinear controller and nonlinear observer systems. The analyzed PEMFC system consists of air compressor motor dynamic equations, air and fuel supply subsystems, a perfect air humidifier and a fuel cell stack. An experimentally validated nonlinear model that reproduces the most typical features of a laboratory PEMFC system is presented. This model is derived based on physics law in stack, including system gases dynamics. The objective of this paper is to introduce a fully analytical model which has been fully validated on a fuel cell system and its auxiliary components. The proposed method can be used as a general modeling guideline for control-oriented purposes. Moreover, it can be successfully implemented in composing a dynamic subsystem, like hydrogen subsystem, as part of the whole nonlinear model.

## 1. Introduction

   Fuel cell is an electrochemical conversion device that converts chemical energy to electrical energy[1]. Of the various fuel cell types, proton exchange membrane fuel cells is much more efficient in terms of generating lower temperature, higher power density, and offering rapid response. However it is not used extensively in real applications due to some drawbacks such as high cost, low reliability, and short lifetime[2, 3]. In particular, PEMFCs are widely used in transportation systems such as cars, buses, and aircrafts. They are suitable for these purposes due to their fast startup time, high power density and favorable power-to-weight

- - - - - - - - - - - - - - - - - - - - - - - - - - - - - - - - - - - - - - - - - - - - - - - - - - - - - -

ratio [4, 5]. However, designing a proper controller under large disturbances is considered as a main problem in a PEMFC. Indeed, an accurate nonlinear model is necessarily required in order to design an advanced controller, under system nonlinearity and uncertainty. Collectively, PEMFC is firstly modeled and then a suitable nonlinear control scheme is applied on its model. In recent years, researches in this area were mostly about control-oriented PEMFC modeling in order to develop advanced control strategies, which can improve the efficiency of fuel cell based systems. Therefore, a nonlinear dynamic model of the fuel cell stack and its auxiliary components is needed for control-oriented researches in this area. Although stack model will be more complicated when involving nonlinear auxiliary components, its quality will be enhanced for control-oriented purposes. Different types of fuel cell models are proposed since last decade. Analytical and non-analytical models are two types of these models.

In analytical and mathematical modeling [6, 7], fuel cell voltage is a nonlinear function of current, stack temperature, the partial pressures of oxygen gas and the partial pressures of hydrogen gas inside cell. In addition, output voltage is the difference between the cell operation voltage and voltage losses. In standard conditions, voltage losses are ohmic, activation and concentration losses. In recent years, an increasing attention has been dedicated to fuel cell simulation and modeling. Nehrir and his colleagues studied steady state response of a hydrogen fuel cell using a transient model [6, 7]. Feliachi and Co-worker also proposed a nonlinear transient model for surveying the dynamic behavior of a solid oxide fuel cell. In fact, they did the investigation considering the effect of rapid load variations on output voltage[8, 9]. Wang and Nehrir in[6, 7] and Sedghisigarchi and Feliachi in [8, 9] used analytical model for analysis, and fuel cell is regarded as a distributed generation system. In addition, the analytical model is applicable in power engineering. Generally, these modeling methods are a kind of mathematical modeling and they are not suitable for PEM fuel cell controller design, especially in complicated conditions. Indeed, an efficient state space model of PEM is required for this purpose.

In 2004, Chiu introduced a small signal model for PEM fuel cell [10]. A linear model is attained by linearization around equilibrium point. Then a controller is designed as an initial step to improve its transient behavior. This controller improved dynamic properties of PEM to some extent. Purkrushpan *et al.* [11-14] developed a detailed mathematical fuel cell stack model that involves nine state variables. This model is resulted considering electrochemical, thermodynamic, and zero-dimensional fluid flow principles. Due to the existence of nonlinear relationship between stack voltage, load current (as polarization curve) and state equations, a nonlinear model is required to design a nonlinear controller. However developing a nonlinear model that can meet these nonlinear relations is a great challenge. In another study, Pukrushpan described

- - - - - - - - - - - - - - - - - - - - - - - - - - - - - - - - - - - - - - - - - - - - - - - - - - - - - -

linear models for fuel cell [12-16]. These models were derived from Jacobian linearization and Taylor expansion methods around equilibrium point. Although the procedure provides a high quality model, its dynamic behavior was not satisfactory against massive load variations. Due to the existence of parametric uncertainties such as parametric coefficients of each cell on kinetic, thermodynamic, and electrochemical foundations, and resistivity of electron flow membrane and external disturbances, linear control approach is not appropriate for PEMFCs. In fact, providing a nonlinear model for the fuel cell is surely a wise decision. In this way, a nonlinear controller may guarantee robust performance of the system against existing disturbances. A nonlinear model of SISO fuel cell stack is studied in [17, 18]. Water, oxygen, and hydrogen pressures are defined as state variables in this research. However, there are a wide variety of models for fuel cell, only a few numbers of them are suitable for control purposes. In [19], a nonlinear dynamic model of a multi input-single output (MISO) fuel cell is described for applying nonlinear control strategy on it.

Moreover, applying nonlinear controller using feedback linearization approach leads to satisfactory transient response and load current against large disturbances. In [20], Kunusch presented a control oriented electrochemical static model of a proton exchange membrane fuel cell. This model includes not only stack voltage and voltage losses, but also some theoretical considerations and semi-empirical analysis based on the experimental data. He also proposed a dynamic model which is more suitable for power generation systems.In [11], only a three-state air supply subsystem is explained. The return manifold, and fuel cell stack are not included in this model. However in Matraji's new approach [21, 22], a four-state air supply subsystem is explained. This model contains some state variables of the stack such as oxygen partial pressure and nitrogen partial pressure. Although it does not include return manifold and oxygen mass of air supply.

## 2. Contribution

In fact, modeling PEMFC systems is difficult because of the existence of subsystems and interactions between differential equations. Introduced models in literature survey only serve analytical purposes, and they are not applicable in control applications due to complicated and time-consuming computations. This research firstly focuses on obtaining a preliminary structure based on physical laws of the system. A nonlinear model of a PEM fuel cell stack in combination with some auxiliary equipment's is then developed. Auxiliaries are compressor, air supply manifold, return manifold, cathode humidifier, line heaters and a single-cell fuel cell stack (Fig. 1). A control-oriented model is proposed in this study. This model is validated experimentally and includes features of a PEMFC stack with its auxiliary components. The proposed representation is a $6^{th}$ order model of PEMFC system in which the measured outputs are the compressor angular speed and the supply and return manifold pressures. Load current is the measured disturbance and the compressor motor voltage is the adjustable input. In this paper, a nonlinear state space model for the fuel cell stack with its auxiliary components is represented. This model is fully convenient for control purposes so that it involves dynamic properties of the system. The presented methods can be used as a general modeling guideline for control-oriented purposes, possible to be generalized to fuel-cell-based systems with similar characteristics.

## 3. PEM Fuel Cells

The reactions for a PEM fuel cell fed with hydrogen in the anode side and oxygen in the cathode side are[1]:

**Anode:** $H_2 \rightarrow 2H^+ + 2e^-$ (1)

**Cathode:** $\frac{1}{2}O_2 + 2H^+ + 2e^- \rightarrow H_2O$ (2)

**Overall reaction:** $H_2 + \frac{1}{2}O_2 \rightarrow H_2O$ (3)

### 3.1. Stack Voltage

**3.1.1. Thermodynamic Voltage:** The maximum amount of electrical energy generated in a fuel cell corresponds to the Gibbs free energy $\Delta G$ of the above reaction. The theoretical potential of a single cell PEM fuel cell generating n electrons can be stated in the following equation:

$$E_0 = \frac{-\Delta G}{nF}$$ (4)

The reversible voltage can also be calculated as follows:

$$E_0 = \frac{-\Delta G}{nF} = \frac{-237340}{2 \times 96485} \frac{J \ mol^{-1}}{As \ mol^{-1}} = 1.229 V$$ (5)

The ideal potential of a fuel cell is 1.229V. The actual fuel cell potential is decreased from its equilibrium point because of irreversible losses.

$$E = 1.229 - 8.5 \times 10^{-4} (T_{fc} - 298.15) + 4.3085 \times 10^{-5} T_{fc} \left[ \ln(P_{H_2}) + \frac{1}{2} \ln(P_{O_2}) \right]$$ (6)

where *E* is the open circuit voltage (Nerst voltage) of a single cell at 25ºC and atmospheric pressure in expanded form as [23],[24, 25].

- - - - - - - - - - - - - - - - - - - - - - - - - - - - - - - - - - - - - - - - - - - - - - - - - - - - - - - - - - -

**3.1.2. Activation Losses:** Activation losses is caused by the slow reactions taking place on the surface of the electrodes [12, 13, 24].

$$v_{act} = \frac{RT}{2\alpha F} \ln\left(\frac{i}{i_0}\right) \tag{7}$$

**3.1.3. Ohmic Losses:** This voltage drop is the straight forward resistance to the flow of electrons through the of the electrodes and various interconnections [12, 13, 24].

$$v_{ohm} = I_{fc} R_{ohm} \tag{8}$$

**3.1.4. Concentration Losses:** As the reactant is consumed rapidly at the electrodes by electrochemical reactions, concentration losses will be generated. Oxygen consumption at the fuel cell cathode during its reactions causes a slight reduction in oxygen concentration at cathode. All in all, these losses cause extensive voltage reduction [20,26].

$$v_{conc} = m \exp(n I_{fc}) \tag{9}$$

**3.1.5. The Operating Cell Voltage:** Several sources contribute to irreversible losses in a practical fuel cell. Losses which are often called polarization over voltage, originate from three sources such as activation, ohmic and concentration losses [7, 26].

These losses result in a cell voltage less than its ideal potential:

$$v_{fc} = E - losses \tag{10}$$

The combined effect of thermodynamics, mass transport kinetics and ohmic resistance determines the cell output voltage as[12, 13, 24]:

$$v_{cell} = E - v_{act} - v_{ohm} - v_{conc} \tag{11}$$

A fuel cell stack consists of several cells in series, to increase the overall voltage of fuel cell. In the following equation, $N$ is the number of cells in series. Fuel cell stack voltage was described by:

$$v_{stack} = N v_{cell} \tag{12}$$

The parametric equations for the single cell voltage and losses of activation, internal resistance and concentration are shown in the following table:

**Table 1:** Fuel cell voltage parameters

| | |
|---|---|
| $N$ | Number of cells |
| $E_0$ | Reversible cell voltage (1.229V) |
| $v_{fc}$ | Fuel cell output voltage |
| $T_{fc}$ | Temperature of fuel cell (K) |
| $T_{atm}$ | Operating cell temperature ( 298.15 K) |
| $P_{H_2}, P_{O_2}$ | Partial pressure of each gas inside |
| $R$ | Universal gas constant (8.4413 J/mole. K) |
| $F$ | Faraday's constant (96439 C/mole) |
| $\alpha$ | Charge transfer coefficient |
| $i$ | Output current density |
| $i_0$ | Exchange current density |
| $I_{fc}$ | Output current |
| $R_{ohm}$ | Area-specific resistance |
| $m \ \& \ n$ | Constants in the mass transfer voltage |

### 3.2. State Space Representation of PEMFC System with Auxiliary Components

In this section, a nonlinear state space model, convenient for fuel cell control, is represented. Dynamic characteristics are well considered in this model. Generally, previous models were not perfect due to the mentioned deficiencies. In this paper, a more efficient model of fuel cell stack and its auxiliary components is implemented. This nonlinear model is extracted from Pukrushpan[12-14] proposed model, and applied as a novel approach. This model includes the following subsystems:

-Air compressor model

-Supply manifold model

-Return manifold model

-Stack model

The following assumptions are also considered for this analysis. Some assumptions are defined in order to attain an accurate PEMFC model including its stack, air supply manifold, return manifold, and the compressor model. Assumptions are stated as follows:

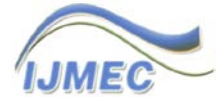

- - - - - - - - - - - - - - - - - - - - - - - - - - - - - - - - - - - - - - - - - - - - - - - - - - - - - - - - - - - -

**1**: The anode pressure is kept constant. This condition can be met by applying a proper control system. Moreover, this assumption is admissible considering the fact that anode pressure varies more slowly than the other state dynamics.

**2**: According to this assumption, temperature and humidity of the air at the inlet of the FC stack are constant. This is actually justifiable regarding very slow variations. Due to their slow time response, a traditional controller can control those states.

**3**: Compressor driving DC motor is modeled regardless of its electrical behavior. It is assumed due to the fact that nominal values of the motor winding impedance include a small time constant. This dynamic is fast and will be soon vanished even if it is considered. Besides, the effect of this dynamic can be derived using small signal analysis if it is needed.

A nonlinear schematic diagram of a PEM fuel cell stack together with some of its auxiliaries is depicted in Fig. 1. The model is derived based on the electrochemical, thermodynamic and fluid flow principles. A $9^{th}$ order nonlinear model of the fuel cell system which is already proposed by Pukrushpan[12-14] is modified. The $9^{th}$ order nonlinear Pukrushpan's model involves air and hydrogen supply sub-systems. In this model the relation between some of the system states are modeled using a lookup table data. In fact, replacement of the lookup table data is a novelty and modification in the model, since it would be possible to determine the intermediate data using these continuous equations without using interpolation method. Meanwhile, in the previous model a static compressor map was used to determine the air flow at the output of the compressor. However, in this paper, the static compressor map is replaced by smooth functions. The latter copes with shortcoming of the switched piece-wise functions which is non-differentiable. Briefly, in the proposed model, smooth functions *e.g.* Compressor model are used instead of piece-wise functions and lookup table. Under above assumptions, the model order reduces from 9 to 6, considering the state vector $x = \begin{bmatrix} \omega_{cp} & P_{sm} & m_{sm} & m_{O2} & m_{N2} & P_{rm} \end{bmatrix}^T$. Using the following definitions:

**Table 2:** Fuel cell variables definition

| Variable / Unit | Meaning |
|---|---|
| $x_1 = \omega_{cp}\ [rad/\sec]$ | Angular motor speed |
| $x_2 = P_{sm}\ [atm]$ | Supply manifold pressure |
| $x_3 = m_{sm} \,[kg]$ | Air mass in the supply manifold |
| $x_4 = m_{O2} \,[kg]$ | Oxygen mass at cathode side |
| $x_5 = m_{N2} \,[kg]$ | Nitrogen mass at cathode side |
| $x_6 = P_{rm} \,[atm]$ | Return manifold pressure |

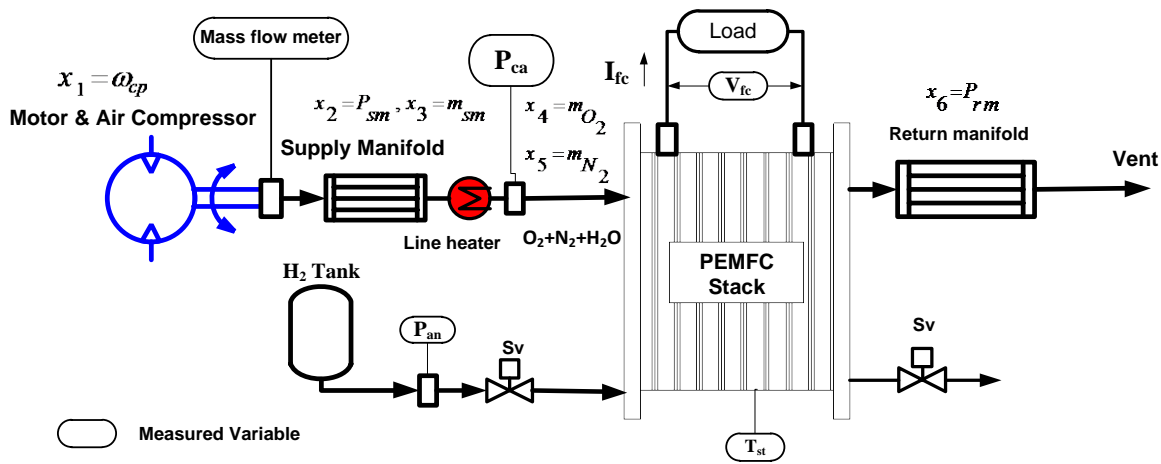

**Figure 1:** Fuel cell and its auxiliary components

### 3.2.1. Compressor Dynamics Model:

The angular speed ($\omega_{cp}$) dynamic is related to the compressor motor torque ($\tau_{cm}$), and required compressor torque ($\tau_{cp}$) by the following equation [12-14, 16]:

$$J_{cp} \frac{d\omega_{cp}}{dt} = \tau_{cm} - \tau_{cp} \qquad (13)$$

where $J_{cp}$ is the compressor motor inertia, $\tau_{cm}(v_{cm}, \omega_{cp})$ is the accelerating torque provided by the motor, and $\tau_{cp}$ is the load torque. The compressor motor torque is calculated through the following static motor equation [13, 14, 27].

$$\tau_{cm} = \eta_{cm} \frac{k_t}{R_{cm}} (v_{cm} - k_v \omega_{cp}) \qquad (14)$$

where $k_t$, $R_{cm}$, and $k_v$ are motor constants and $\eta_{cm}$ is the motor mechanical efficiency.

The load torque required to drive the compressor is also determined by Eq. (15) [28, 29].

$$\tau_{cp} = \frac{\pi}{30}(\alpha_0 + \alpha_1 x_1 + \alpha_{00} + \alpha_{10} x_1 + \alpha_{20}(x_1)^2 + \alpha_{01} x_2 + \alpha_{11} x_2 x_1 + \alpha_{02}(x_2)^2)) \quad (15)$$

where the values of the $\alpha_{ij}$ constants are [28, 29]:

$$\alpha_{00} = 0, \alpha_{10} = 0.0058, \alpha_{20} = -0.0013, \alpha_{01} = 3.25 \times 10^{-6}, \alpha_{11} = -2.80 \times 10^{-6}, \alpha_{02} = -1.37 \times 10^{-9},$$
$$\alpha_1 = 3.92 \times 10^{-6}, \alpha_0 = 4.1 \times 10^{-4}. \quad (16)$$

A static compressor map was attained in the previous studies by curve fitting [12-15]. This map and piecewise continuous functions were used to determine the outlet air flow rate through the compressor [28, 29]. In this study, differentiable continuous polynomial functions are used instead of static compressor map. Therefore, the compressor air flow rate is given by Eq. (17).

$$W_{cp} = \beta_{00} + \beta_{10}.(P_{sm}) + \beta_{20}.(P_{sm})^2 + \beta_{01}.(\omega_{cp}) + \beta_{11}.P_{sm}.\omega_{cp} + \beta_{02}.(\omega_{cp})^2 \quad (17)$$

The values of the $B_{ij}$ constants are [29] :

$$\beta_{00} = 4.83 \times 10^{-5}, \beta_{10} = -5.42 \times 10^{-5}, \beta_{20} = 8.79 \times 10^{-6}, \beta_{01} = 3.49 \times 10^{-7},$$
$$\beta_{11} = 3.55 \times 10^{-13}, \beta_{02} = -4.11 \times 10^{-10}. \quad (18)$$

By substituting Eqs. (13) to (17), state equation for the rotational speed of the compressor is represented by Eq. (19).

$$\dot{x}_1 = \tau_{cm} - \tau_{cp} \Rightarrow \dot{x}_1 = \frac{\eta_{cm}}{J_{cp}} \frac{k_t}{R_{cm}} (v_{cm} - k_v x_1) - \frac{\tau_{cp}}{J_{cp}} \quad (19)$$

According to Eqs. (14) and (19), the angular compressor speed ($\omega_{cp}$), can be controlled by the compressor motor voltage ($v_{cm}$).

### 3.2.2. Supply manifold air pressure and air mass dynamics model

The cathode supply manifold includes pipes and stack manifold volume between compressor and fuel cell stack. The rate of mass variations in supply manifold is determined by the mass conservation principle, and the rate of pressure variations in supply manifold is governed by energy conservation law. Air pressure and air mass in the supply manifold is given by differential Eqs. (20) and (21). These differential equations are related to the compressor air flow ($W_{cp}$), and the temperature of compressor exit flow ($T_{cp}$) [12-15].

$$\dot{x}_2 = \frac{\gamma R_a}{M_a^{atm} V_{sm}} (W_{cp} T_{cp} - W_{sm} T_{sm}) \quad (20)$$

$$\dot{x}_3 = W_{cp} - W_{sm} \quad (21)$$

where $V_{sm}$ denotes the supply manifold volume. The air temperature in the supply manifold ($T_{sm} = \frac{P_{sm} V_{sm} M_a^{atm}}{R \, m_{sm}}$) is also related to the two states ($x_2$) and ($x_3$).

$$T_{sm} = \frac{x_2 V_{sm} M_a^{atm}}{R \, x_3} \quad (22)$$

According to Eq. (21), $x_3 = m_{sm} \, [kg]$ denotes the air mass in the supply manifold. This parameter is associated with the compressor exit flow ($W_{cp}$), and the supply manifold exit flow. Eq. (20) also represents $x_2 = P_{sm} \, [atm]$ as the supply manifold pressure; this state is related to the gas temperature in supply manifold ($T_{sm}$). $T_{cp}$ is defined as the temperature of the gas leaving the compressor, and it is given by:

$$T_{cp} = T_{atm} + \frac{T_{atm}}{\eta_{cp}} \left[ \left( \frac{P_{sm}}{P_{atm}} \right)^{\frac{\gamma}{\gamma-1}} - 1 \right] \quad (23)$$

The rate of supply manifold outlet air ($W_{sm}$) is stated as a function of $P_{sm}$ and $P_{ca}$. $W_{sm}$ is obtained via a linearized nozzle equation, then the following equations can be derived [12-15].

$$W_{sm} = K_{sm,out} (P_{sm} - P_{ca}) \quad (24)$$

The cathode pressure is calculated as the summation of three components (oxygen, nitrogen and vapor) using Dalton's law of partial pressures.

$$P_{ca} = P_{v,ca} + P_{O_2} + P_{N_2} = P_{v,ca} + c_1 x_5 + c_2 x_4 \quad (25)$$

Where the two parameters $c_2$ and $c_2$ are as follows:

$$c_2 = \frac{R_{O_2} T_{st}}{V_{ca}}, \quad c_1 = \frac{R_{N_2} T_{st}}{V_{ca}} \tag{26}$$

Substitution of Eqs. (22) to (25) into Eq. (20) results in the following equation of $x_2$.

$$\dot{x}_2 = \frac{\gamma R_a}{V_{sm}} (-K_{sm,out} x_2 + K_{sm,out} P_{v,ca} + K_{sm,out} \frac{x_5}{M_{N_2}} c_1 + K_{sm,out} \frac{x_4}{M_{O_2}} c_2) \frac{\gamma x_2}{x_3}$$
$$+ W_{cp} (T_{atm} + \frac{T_{atm}}{\eta_{cp}} ((\frac{x_2}{P_{atm}})^{\frac{\gamma-1}{\gamma}} - 1)) \tag{27}$$

By substituting Eq. (24) into Eq. (21), the following equation for $x_3$ is derived.

$$\dot{x}_3 = W_{cp} - K_{sm,out} x_2 + K_{sm,out} P_{v,ca} + K_{sm,out} \frac{x_5}{M_{N_2}} c_1 + K_{sm,out} \frac{x_4}{M_{O_2}} c_2 \tag{28}$$

**3.2.3 Cathode Flow Model:** Several assumptions are considered for developing the cathode flow model [12-15].

- All gases obey the ideal gas law
- The air temperature inside cathode is assumed to be equal to the stack temperature.
- Some variables such as pressure, temperature and humidity of the exiting flow are assumed to be the same as the variables inside cathode.
- The flow channel and cathode backing layer are lumped into one cathode volume and thus the minute variations are ignored.

Based on the mass conservation principle, mass equations of oxygen and nitrogen gases inside the cathode are given by Eq.(29) and Eq. (30) [16, 30]:

$$\dot{x}_4 = W_{O_2,in} - W_{O_2,out} - W_{O_2,reacted} \tag{29}$$

$$\dot{x}_5 = W_{N_2,in} - W_{N_2,out} \tag{30}$$

Mass conservation yields governing equations for oxygen, nitrogen inside the cathode volume given by $W_{N_2,in}$ and $W_{N_2,out}$.

Thus, the oxygen variables of $W_{O_2,out}$, $W_{O_2,reacted}$ and $W_{O_2,in}$, and nitrogen variables are required for

- - - - - - - - - - - - - - - - - - - - - - - - - - - - - - - - - - - - - - - - - - - - - - - - - - - - - - -

calculating the oxygen mass state variable ($x_4 = m_{O2} \, [kg]$) and nitrogen mass state variable ($x_5 = m_{N2} \, [kg]$).

So Eq. (31) shows the oxygen and nitrogen flow rate into the stack from the supply manifold ($W_{sm,out}$).

$$W_{N_2,in} = y_{N_2} \frac{1}{1+\Omega_{atm}} W_{sm,out} \quad , \quad W_{O_2,in} = y_{O_2} \frac{1}{1+\Omega_{atm}} W_{sm,out} \tag{31}$$

Outlet flow rate of supply manifold is related to the difference between the pressures of the upstream and downstream gases, as follows:

$$W_{sm,out} = k_{sm,out} (P_{sm} - P_{ca}) \tag{32}$$

Mass fraction of oxygen and nitrogen in dry atmospheric air are denoted as $y_{O_2} = X_{O_2} \frac{M_{O_2}}{M_a^{atm}}$ and $y_{N_2} = (1 - X_{O_2}) \frac{M_{N_2}}{M_a^{atm}}$ respectively. The oxygen mole fraction in dry air is $M_a^{atm} = X_{O_2} M_{O_2} + (1 - X_{O_2}) M_{N_2}$ and oxygen mass fraction in dry atmospheric air is $X_{O_2} = 0.21$.

The humidity ratio of cathode inlet and atmospheric (at compressor inlet) can be defined through Eq. (33):

$$\Omega_{ca,in} = \frac{M_v}{M_a} \frac{\phi_{ca,in}^{des} P_{sat}^{st}}{P_{sm}\left(1 - \phi_{atm} \frac{P_{sat}^{atm}}{P_{atm}}\right)} \quad , \quad \Omega_{atm} = \frac{M_v}{M_a} \frac{\phi_{atm} \frac{P_{sat}^{atm}}{P_{atm}}}{1 - \phi_{atm} \frac{P_{sat}^{atm}}{P_{atm}}} \tag{33}$$

$P_{sat}^{st}$ represents the vapor saturation pressure, and relative humidity is defined as $\phi_{ca} = \min(1, \frac{m_{v,ca} R T_{st}}{P_{sat}^{st} M_v V_{ca}})$.

The mass flow rate of oxygen and nitrogen exiting the cathode is relevant to the oxygen mass, nitrogen mass, total mass of cathode gas ($m_{ca}$), and the flow rate of exiting mass of cathode ($W_{ca,out}$). Flow rate of the exiting oxygen and nitrogen mass is calculated through Eq. (27) [12-14, 30].

$$W_{O_2,out} = \frac{m_{O_2}}{m_{ca}} (W_{ca,out}) \tag{34}$$

$$W_{N_2,out} = \frac{m_{N_2}}{m_{ca}} (W_{ca,out}) \tag{35}$$

Regarding Eq. (36), a linearized nozzle equation is used to calculate the cathode exit flow rate. It is shown that exit flow of cathode is related to the difference between the cathode input pressure and cathode output pressure [31].

$$W_{ca,out} = k_{ca,out}(P_{ca} - P_{rm}) \tag{36}$$

The total mass of the cathode gas, includes oxygen mass, nitrogen mass and vapor mass
($m_{ca} = m_{O_2} + m_{N_2} + \frac{V_{ca} P_{va,ca} M_v}{R T_{st}}$)[14].

The oxygen reaction rate ($W_{O_2,reacted}$) in cathode is related to the stack current $I_{st}$, and can be calculated by the following electrochemical equation [14, 30, 31]:

$$W_{O_2,reacted} = M_{O_2} \frac{n I_{fc}}{4F} \tag{37}$$

### 3.2.3. Return Manifold Model:
The return manifold pressure is governed by the mass conservation and the ideal gas law as[29, 31]:

$$\dot{x}_6 = \frac{R_a T_{rm}}{V_{rm}} (W_{ca,out} - W_{rm,out}) \tag{38}$$

where $V_{rm}$ denotes the return manifold volume.

Return manifold temperature is also assumed equal to stack temperature ($T_{st} = T_{rm}$).

$W_{ca,out}$ is required for calculating the return manifold pressure ($x_6 = P_{rm}$ [atm]). Output flow of return manifold can be calculated through the following polynomial equation [31-33].

$$W_{rm,out} = \sum_{i=0}^{5} P_{a_i} P_{rm}^i \tag{39}$$

where the values of $P_{a_i}$ are[28, 31, 32]:

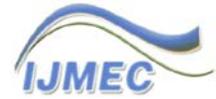

- - - - - - - - - - - - - - - - - - - - - - - - - - - - - - - - - - - - - - - - - - - - - - - - - - - - - -

$$P_{a_0} = 1.248 \cdot 10^{-3}, P_{a_1} = -1.96 \cdot 10^{-3}, P_{a_2} = -1.52 \cdot 10^{-3}, P_{a_3} = -2.12 \cdot 10^{-3}, P_{a_4} = -27.7 \cdot 10^{-3},$$
$$P_{a_3} = -78 \cdot 10^{-3}. \qquad (40)$$

In order to determine the $W_{rm,out}$ value in this equation, data is collected by implementing several experiments in different equilibrium points [28, 32]. The final equation for $x_6$ is stated as follows:

$$\dot{x}_6 = \frac{R_a T_{rm}}{V_{rm}}\left( K_{ca,out}\left(\frac{x_4}{M_{O_2}} c_2 + \frac{x_5}{M_{N_2}} c_1 + P_{v,ca} - x_6\right) - \left(\sum_{i=1}^{5} P_{a_i} x_6^i\right)\right) \qquad (41)$$

**Table 3:** PEMFC parameters

| Parameter | Symbol / SI Unit | Value |
|---|---|---|
| Motor torque constant | $K_t$ $(Nm/A)$ | 0.0153 |
| Motor winding resistance | $R_{cm}$ $(ohm)$ | 0.82 |
| Motor back-emf | $K_v$ $(V/rad/sec)$ | 0.0153 |
| Compressor efficiency | $\eta_{cp}$ | 0.8 |
| Motor mechanical efficiency | $\eta_{cm}$ | 0.98 |
| Supply manifold volume | $V_{sm}$ $(m^3)$ | 0.02 |
| Single stack cathode volume | $V_{ca}$ $(m^3)$ | 0.005 |
| Return manifold volume | $V_{rm}$ $(m^3)$ | 0.005 |
| Supply manifold outlet orifice constant | $K_{sm,out}$ $(kg/sec/Pa)$ | $0.3629 \times 10^{-5}$ |
| Cathode outlet orifice constant | $K_{ca,out}$ $(kg/sec/Pa)$ | $0.2177 \times 10^{-5}$ |
| Compressor diameter | $dc$ $(m)$ | 0.2286 |
| Compressor and motor inertia | $J_{cp}$ $(kg\, m^2)$ | $5 \times 10^{-5}$ |
| Oxygen mole fraction at cathode inlet | $y_{O_2,in}$ | 0.21 |

- - - - - - - - - - - - - - - - - - - - - - - - - - - - - - - - - - - - - - - - - - - - - - - - - - - - - - -

**Table 4:** Physical parameters in the modeling of PEMFC

| Parameter | Symbol / SI Unit | Value |
|---|---|---|
| Atmospheric pressure | $P_{atm}$ (Pa) | 101325 |
| Average ambient air relative humidity | $\Phi_{atm}$ | 0.5 |
| Saturation pressure | $P_{sat,Tatm}$ (Pa) | $3.1404 \times 10^3$ |
| Air-specific heat ratio | $\gamma$ | 1.4 |
| Air density | $C_p$ (J/kg/K) | 1004 |
| Air gas constant | $R_a$ (J/mol/K) | 286.9 |
| Oxygen gas constant | $R_{O_2}$ (J/kg/K) | 259.8 |
| Nitrogen gas constant | $R_{N_2}$ (J/kg/K) | 296.8 |
| Vapor gas constant | $R_v$ (J/kg/K) | 461.5 |
| Molar mass of air | $M_a$ (kg/mol) | $28.97 \times 10^{-3}$ |
| Molar mass of oxygen | $M_{O_2}$ (kg/mol) | $32 \times 10^{-3}$ |
| Molar mass of nitrogen | $M_{N_2}$ (kg/mol) | $28 \times 10^{-3}$ |
| Molar mass of vapor | $M_v$ (kg/mol) | $18.02 \times 10^{-3}$ |

Considering above equations, it can be seen that the system state equations are complicated since they have many parameters. The unique state-space complete model equations are as follows:

$$\dot{x}_1 = \frac{\eta_{cm}}{J_{cp}} \frac{k_t}{R_{cm}} (v_{cm} - k_v x_1) - \frac{\tau_{cp}}{J_{cp}} \tag{42}$$

$$\dot{x}_2 = \frac{\gamma R_a}{V_{sm}} (-K_{sm,out} x_2 + K_{sm,out} P_{v,ca} + K_{sm,out} \frac{x_5}{M_{N_2}} c_1 + K_{sm,out} \frac{x_4}{M_{O_2}} c_2) \frac{\gamma x_2}{x_3}$$

$$+ W_{cp} (T_{atm} + \frac{T_{atm}}{\eta_{cp}} ((\frac{x_2}{P_{atm}})^{\frac{\gamma-1}{\gamma}} - 1)) \tag{43}$$

$$\dot{x}_3 = W_{cp} - K_{sm,out} x_2 + K_{sm,out} P_{v,ca} + K_{sm,out} \frac{x_5}{M_{N_2}} c_1 + K_{sm,out} \frac{x_4}{M_{O_2}} c_2 \tag{44}$$

$$\dot{x}_4 = -\frac{x_4}{x_4 + x_5 + c_3} K_{ca,out} (-x_6 + P_{v,ca} + \frac{x_5}{M_{N_2}} c_1 + \frac{x_4}{M_{O_2}} c_2) + y_{O_2,in} K_{sm,out}$$
$$(x_2 - \frac{x_4}{M_{O_2}} c_2 - P_{v,ca} - \frac{x_5}{M_{N_2}} c_1) - n \frac{M_{O_2}}{4F} I_{fc}$$
(45)

$$\dot{x}_5 = (1 - X_{O_2,in}) (1 + \Omega_{atm})^{-1} K_{sm,out} (x_2 - \frac{x_4}{M_{O_2}} c_2 - \frac{x_5}{M_{N_2}} c_1 - P_{v,ca}) - \frac{x_5}{x_4 + x_5 + c_3} K_{ca,out}$$
$$(-x_6 + \frac{x_4}{M_{O_2}} c_2 + \frac{x_5}{M_{N_2}} c_1 + P_{v,ca})$$
(46)

$$\dot{x}_6 = \frac{R_a T_{rm}}{V_{rm}} (K_{ca,out} (\frac{x_4}{M_{O_2}} c_2 + \frac{x_5}{M_{N_2}} c_1 + P_{v,ca} - x_6) - (\sum_{i=1}^{5} P_{a_i} x_6^i))$$
(47)

Nonlinear compact form of the state space equations of this model is illustrated in Eq.s (48) and (49).

$$\dot{x} = f(x) + g(x) u + \varphi(x) d \ , \ x = \begin{bmatrix} \omega_{cp} & P_{sm} & m_{sm} & m_{O_2} & m_{N_2} & P_{rm} \end{bmatrix}^T , \begin{bmatrix} u = v_{cm} \end{bmatrix}, \begin{bmatrix} d = I_{fc} \end{bmatrix}$$
(48)

$$f(x) = \begin{bmatrix} f_1(x_1, x_2, W_{ca}(x_1, x_2)) \\ f_2(W_{ca}(x_1, x_2), x_2, x_3, x_4, x_5) \\ f_3(x_1, x_2, x_4, x_5) \\ f_4(x_2, x_4, x_5, x_6) \\ f_5(x_2, x_4, x_5, x_6) \\ f_6(x_4, x_5, x_6) \end{bmatrix}, \ g(x) = \begin{bmatrix} \eta_{cm} \frac{k_t}{J_{cp} R_{cm}} \\ 0 \\ 0 \\ 0 \\ 0 \\ 0 \end{bmatrix}, \ \varphi(x) = \begin{bmatrix} 0 \\ 0 \\ 0 \\ -n \frac{M_{O2}}{4F} \\ 0 \\ 0 \end{bmatrix}$$
(49)

This model is nonlinear, and therefore appropriate for designing nonlinear controller or observer. In fact, there is no need to linearize the model around operating point or use linear controller. Input voltage to compressor motor is used as the main control action and the fuel cell stack current is regarded as the measured disturbance. $x \in \mathbb{R}^6$ are known as model states, and $f \in \mathbb{R}^6 \to \mathbb{R}^6$ is the continuous vector function, representing the system dynamics. $g \in \mathbb{R}^6$ and $\varphi(x) \in \mathbb{R}^6$ are constant vectors, defining input gain matrix and disturbance respectively.

- - - - - - - - - - - - - - - - - - - - - - - - - - - - - - - - - - - - - - - - - - - - - - - - - - - - - - - - -

## 4. Experimental Set Up and Preliminaries

This study mainly focuses on the voltage-current characteristics of a PEM fuel cell. The catalyst content of the cathode is 0.4mg Pt cm$^{-2}$ and the catalyst content of the anode is 0.2 mg Pt cm$^{-2}$. Carbon cloth is used as a gas diffusion layer. The membrane is E-TEK Nafion 112 with total surface area of 64 cm$^2$ and effective surface area of 25 cm$^2$. The thickness of MEA is 2 mm. Bipolar plates are composite graphite plates. Spiral pattern is selected for the current model of cathode and anode sides. Two parallel channels with 1 mm width and 1 mm depth are adapted. Belt width between two channels is 1 mm, and path length is 0.58 m for each channel. A conductive copper plate is used for collecting and conducting the effective current. Fiber plates are used as the insulators between copper plates. They are suitable for this purpose because of efficient thermal stability. Pure hydrogen is used as fuel as well. Volumetric flow controller (OMEGA) is applied to control the reactants flow. In order to humidify the reactant gases, they firstly pass through a hot water tank with a specific temperature and enter the cell. By placing a pressure regulator in both sides of the cell, it can be tested under different pressures. Then the applied current is increased step by step from 1 to 15 A, and cell performance is analyzed in each step.

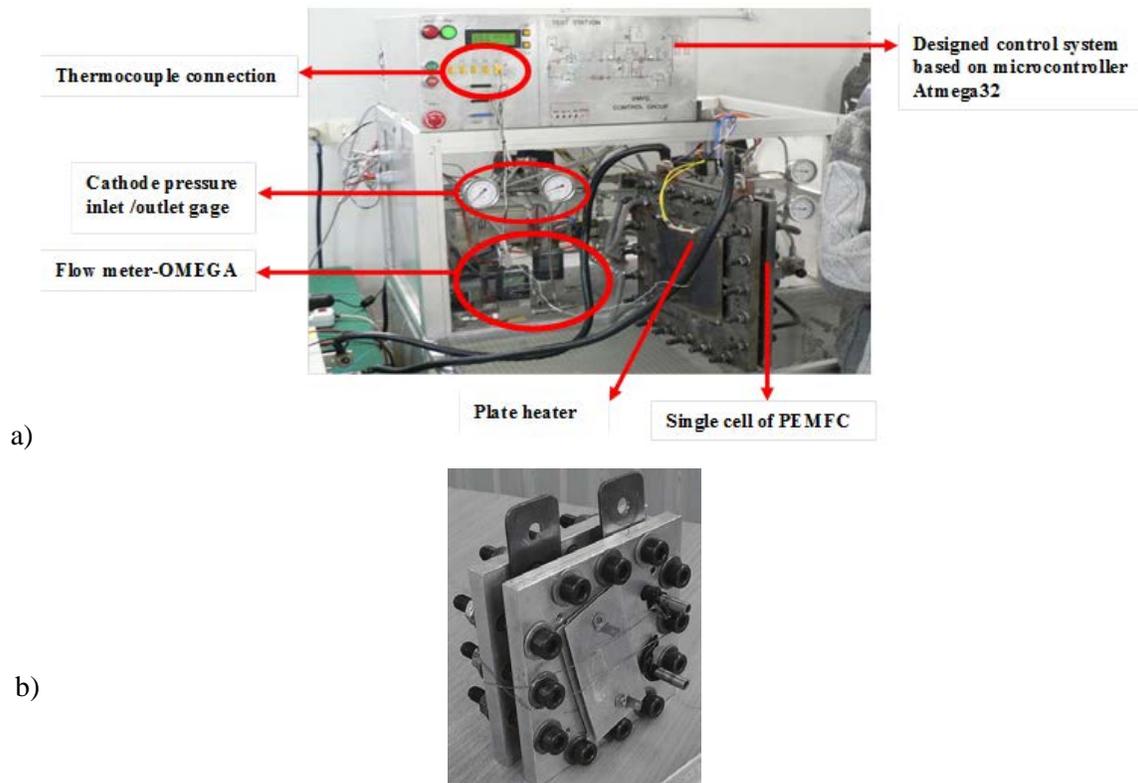

a)

b)

**Figure 2: a):** Experimental PEMFC laboratory and **b):** PEM single cell with an effective area of 25 cm$^2$

- - - - - - - - - - - - - - - - - - - - - - - - - - - - - - - - - - - - - - - - - - - - - - - - - -

### 4.1. Hardware of Designed Control System Based on Microcontroller

Hardware of the control system consists of the following blocks: 1) Central processing unit, 2) Analog inputs, 3) Digital inputs, 4) Analog outputs, 5) Digital outputs, 6) RS232 serial port.

Control system is composed of a main board which relies on an ATmega128 microcontroller with the following practical characteristics:

- Functional frequency of 16 $MH_Z$.
- Eight analog inputs.
- Six PWM outputs.
- Two serial ports.

Main board of the control system is shown in the following figure. The pressure can be measured in the analog input module by controller with the following properties: Analog input, 0-6bars, 4-20mA. The temperature can also be measured by thermocouple type K with the following characteristics: Analog input, 1° C/1mV. In addition, flow meters are measurable through RS232 serial port. Current and voltage of Electronic load Chroma 63201 can be determined through RS232 serial port.

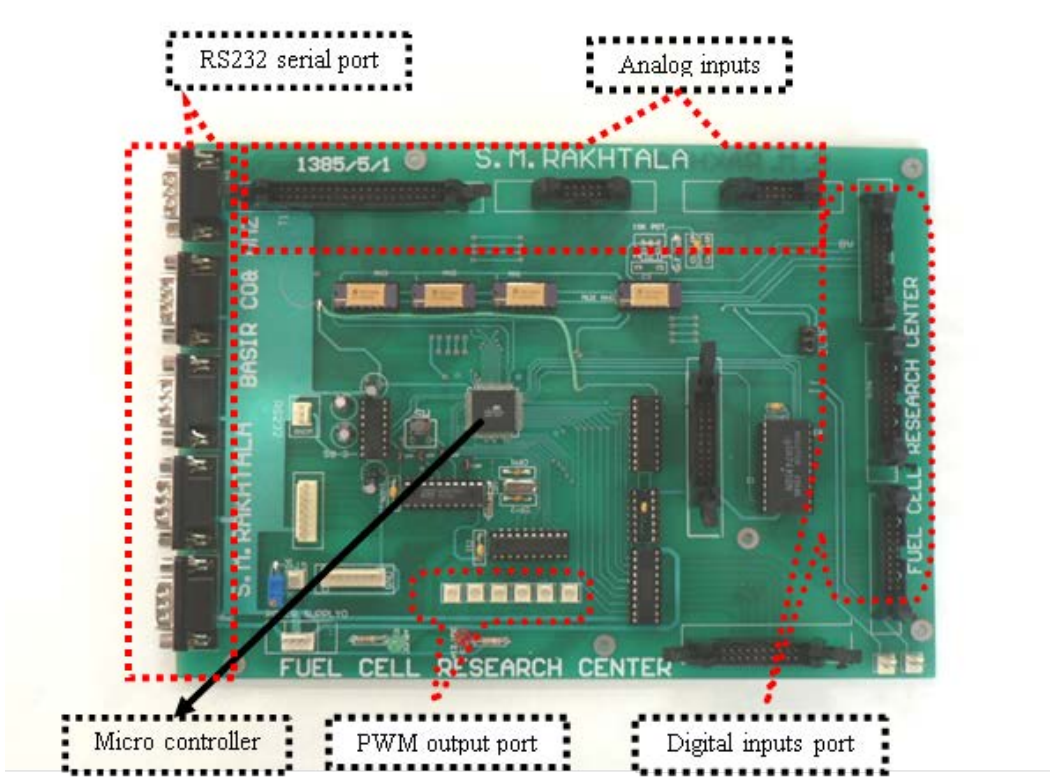

**Figure3 :** Main board of the PEM fuel cell control system

Using the microcontroller, a proper PWM pulse output will be generated on the motor drive board which can control motor speed. Tape heaters are used for heating hydrogen and air. Fuel cell and air temperatures are controlled using a conventional controller derived from feedback of K type thermocouples. Hydrogen and air temperatures are also regulated by the PWM pulse output on heaters which is produced by the microcontroller. Moreover, microcontroller generates proper PWM pulse output on plate heater in order to control stack temperature. The following figure shows block diagram of the control unit, its auxiliary components and their connections.

## 5. Experimental results and discussions for PEMFC

Several experiments were conducted on the above constructed hydrogen fuel cell. The experiments were carried out at various temperatures, pressures and cathode flow rates. The simulation of the nonlinear model is matched with the experimental data.

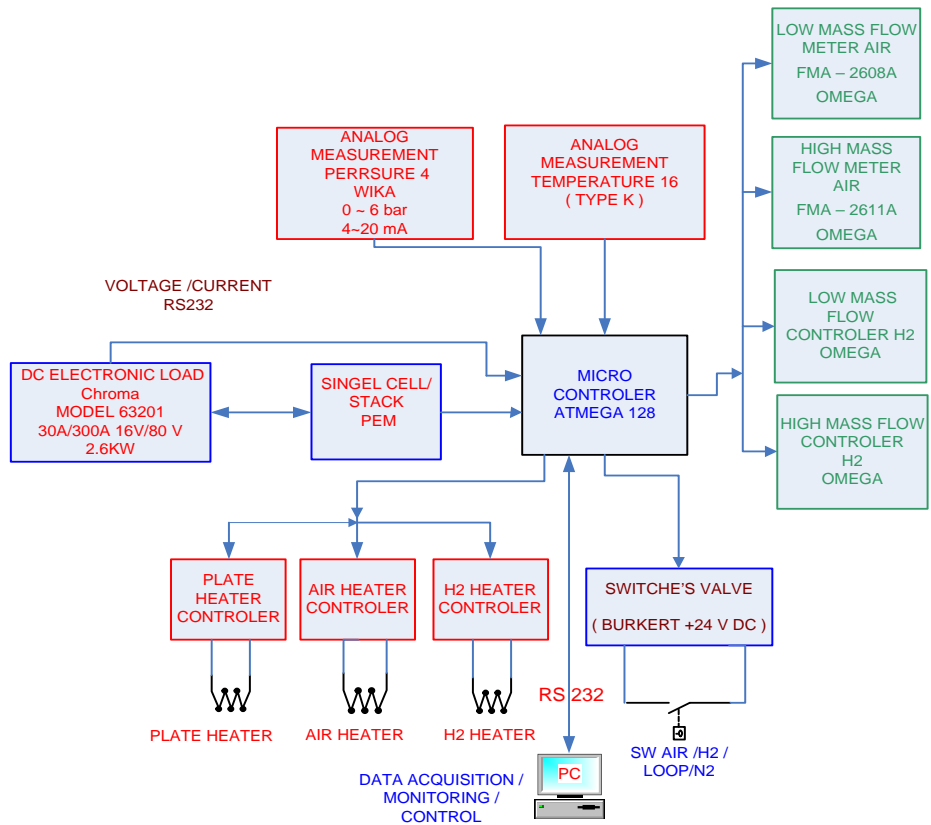

**Figure 4:** Overall block diagram of fuel cell control system.

### 5.1. Air flow rate effect on the polarization curve

The cathode air flow rates (*W*cp) were varied from $0.00333\ kg/s$ $(0.2\ Slpm)$ to $0.00833$ $(0.5\ Slpm)$, and $P_{H_2} = P_{O_2} = 1\ bar$. It is important to emphasize that in all the experiments presented in this section, the air and hydrogen dew point was 60 ºC, while the line heaters and stack temperatures were fixed at 60 ºC in order to avoid water condensation inside the stack. The temperature and humidity of the inlet gases of the fuel cell stack are constant. A PI controller is used to regulate those. Considering these externally regulated temperatures, the relative humidity of the inlet gases was 0.75. The stack current profile was set to vary from 0 A to 12 A, and the anode hydrogen flow was set fixed at $0.00333$Kg/s $(0.2\ Slpm)$. The hydrogen stoichiometry remains above 5 even at the highest current (15A), so the losses due to hydrogen concentration can be neglected. The following figure (Fig. 5) represents the polarization curves obtained at constant pressure (1 bar) and different air flow rates. Fig. 5 shows the effect of changes in air flow rate on the cell performance, which indicates that this effect is very marginal.

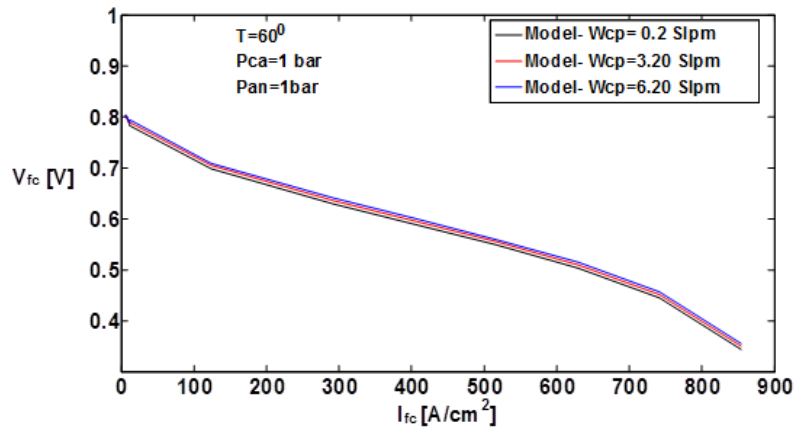

**Figure 5:** Polarization curve of the single cell with model data (constant levels of cathode and anode pressure and different levels of air flow rate).

### 5.2. Effect of pressure on the polarization curve

Fig. 6 depicts the polarization curves obtained at constant air flow of 0.00333 kg/s $(0.2\ Slpm)$ and constant cell temperature of 60 ºC. According to the simulation and experimental results in Fig. 6, it is obvious that the obtained results are satisfactory and reliable. According to Fig. 7, cell voltage increases as working pressure levels are increased from 0.5 to 1.5 bar. Furthermore, higher cathode pressure rate leads to higher cell current and voltage. The simulation results of the dynamic equation shows that Nerst voltage equation and fuel cell voltage equation are matched with the experiment results. The performance of the fuel cell increases with the rise of cell operating pressure, as shown in Fig. 7. An increase of the pressure in both cathode sides increases the reactants concentration in the cell and therefore improves the overall performance.

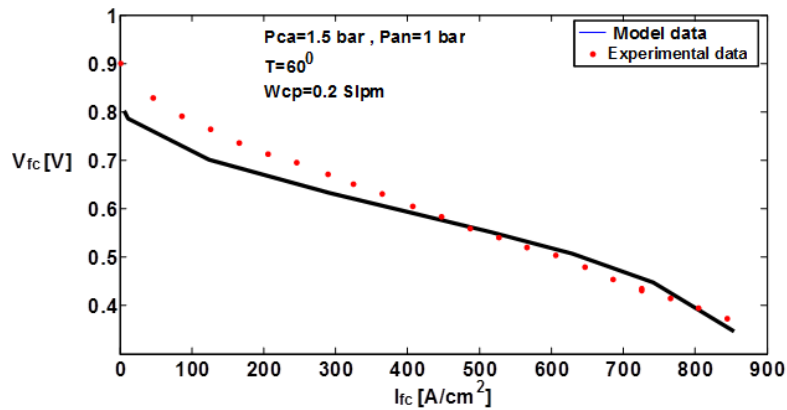

**Figure 6:** Polarization curve of the single cell with experimental and model data (constant air flow rate and 1.5 bar of cathode pressure).

Fig. 8 shows experimental and simulation results at different cathode pressures. These satisfactory results validate PEM fuel cell modeling approach at various cathode pressures.

### 5.3. Temprature effect

Fig.9 shows V–I curve at 55ºC. It is appreciable that the presented state space model derived from Eq.(42-47) successfully predicts fuel cell static V–I behavior under tested conditions at temperature of 55 ºC, air flow rate of 0.2 Slpm, and cathode pressure of 1 bar. Fig. 10 shows cell polarization curve in several temperatures and a comparison of model results with experimental data. Pressure in cathode and anode are 1 bar. Experiments are performed at two different temperatures of 45ºC and 70ºC.

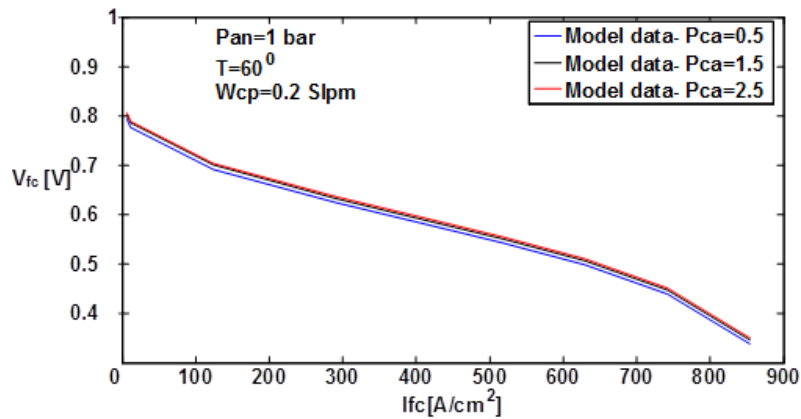

**Figure7:** Polarization curve of the single cell with model data (constant air flow rate and different levels of cathode pressure)

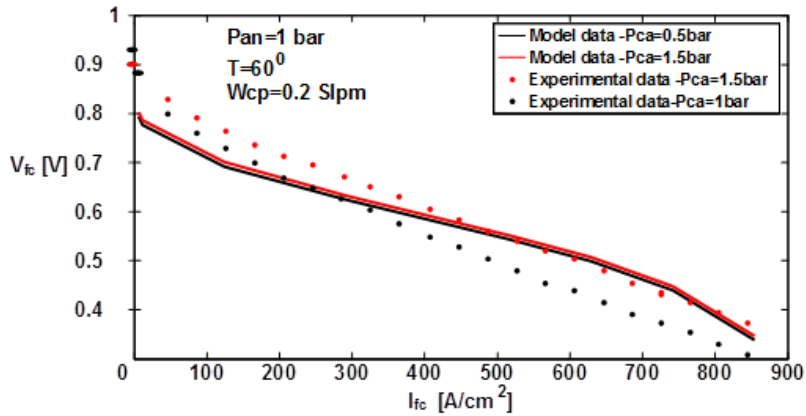

**Figure 8:** Polarization curve of the single cell with experimental and model data (constant air flow rate and different levels of cathode pressure).

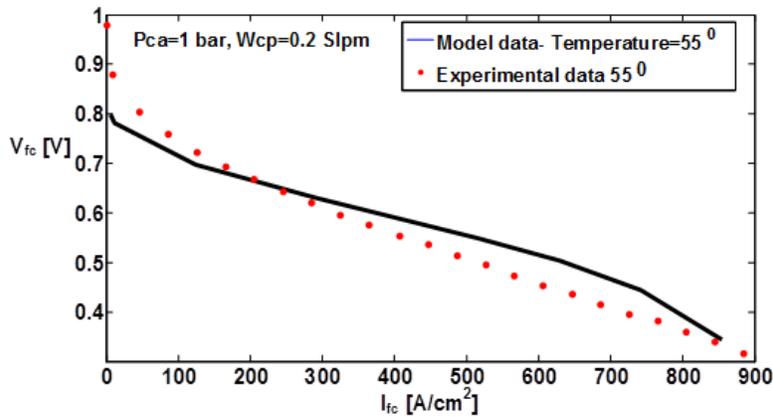

**Figure 9:** The polarization curve of the single cell with experimental and model data at $T_{st} = 55\,^\circ C$

## Conclusion

Fuel cell is a complex system, consists of air compressor motor, air and fuel supply subsystems, air humidifier and a stack. In this survey, a 6$^{th}$ order nonlinear state-space model of a PEM fuel cell system is offered and discussed considering modular subsystems. Model validation is also performed on a single cell PEMFC in the experimental PEMFC laboratory. This model is suitable for nonlinear control purposes and nonlinear observer design. Indeed, the proposed method is a general modeling approach for control design purposes. Moreover, model validation is carried out on the entire operation range of fuel cell-based systems.

- - - - - - - - - - - - - - - - - - - - - - - - - - - - - - - - - - - - - - - - - - - - - - - - - - - - - - - -

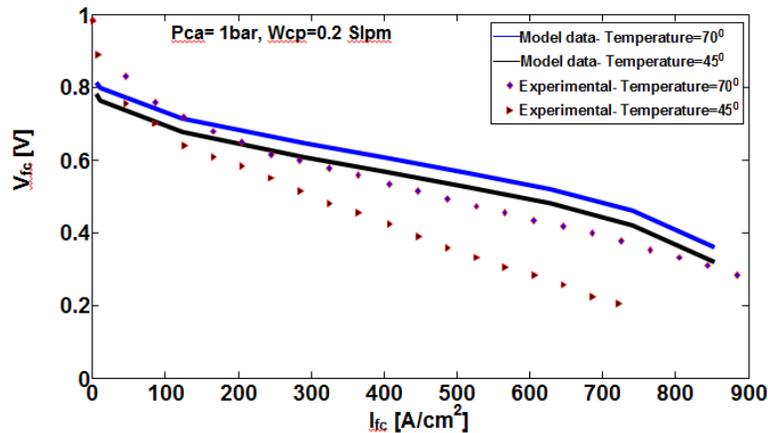

**Figure 10:** The polarization curve of the single cell with experimental and model data at different temperatures

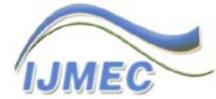